\documentclass[final]{appolb}
\usepackage{epsfig}
\usepackage{psfrag}
\usepackage{amssymb,amsmath}

\begin{document}
\title{Duality between Wilson Loops and Scattering Amplitudes
\thanks{Presented at the 48th Cracow School of Theoretical Physics ``Aspects of Duality'', June 13-22, 2008, Zakopane, Poland}%
}
\author{J. M. Henn
\address{Humboldt-Universit\"at zu Berlin, Institut f\"ur Physik,\\
Newtonstra\ss{}e 15, D-12489 Berlin, Germany\\
henn@physik.hu-berlin.de}
}

\maketitle

\begin{abstract}
We summarise the status of an intriguing new duality between
planar maximally helicity violating scattering amplitudes and light-like Wilson loops in
$\mathcal{N}=4$ super Yang-Mills.
In particular, we focus on the role played by (dual) conformal symmetry,
which is made predictive by deriving anomalous conformal Ward identities for the Wilson
loops. Assuming the duality, the conformal symmetry of the dual Wilson loops 
becomes an unexpected new symmetry of scattering amplitudes in $\mathcal{N}=4$ SYM.
\end{abstract}
\par {HU-EP-08/46}
\par {LAPTH-CONF-1284/08}

\section{Introduction}
We will discuss planar maximally helicity violating (MHV) scattering amplitudes in the maximally supersymmetric Yang-Mills theory in four dimensions, $\mathcal{N}=4$ SYM.
There are many reasons for being interested in scattering amplitudes in $\mathcal{N}=4$ SYM,
ranging from practical applications like the computation of similar amplitudes in QCD to more theoretical motivations.\\

One motivation comes from the fact that from the infrared divergent part of scattering amplitudes 
one can compute the cusp anomalous dimension $\Gamma_{\rm cusp}$ \cite{Polyakov:1980ca,Korchemsky:1985xj,Korchemsky:1987wg}. The latter has received considerable
attention over the last years in the study of the AdS/CFT
correspondence. Its value is predicted (in principle at any given
order) from conjectured integrable models that describe the
spectrum of anomalous dimensions in $\mathcal{N}=4$ SYM \cite{Beisert:2006ez}. 
Therefore, knowing $\Gamma_{\rm cusp}$ to high orders in perturbation theory
is important to test and fine-tune these models. The three- and
four-loop values of $\Gamma_{\rm cusp}$ were indeed determined from
four-gluon scattering amplitudes \cite{Bern:2005iz,Bern:2006ew,Cachazo:2006az}.\\

However, the scattering amplitudes themselves also reveal interesting
properties, on which I will focus in this talk. 
An iterative structure for (planar) MHV scattering amplitudes in 
$\mathcal{N}=4$ SYM was uncovered by Anastasiou,
Bern, Dixon and Kosower (ABDK) \cite{Anastasiou:2003kj} and
generalised to higher loops by Bern, Dixon and Smirnov (BDS)
\cite{Bern:2005iz}. In particular, it turns out that the finite
part of the scattering amplitudes seems to be much simpler than 
could be expected on general grounds.\\

We will argue that a possible explanation for this surprising simplicity
is a new symmetry of scattering amplitudes, dual conformal symmetry. This is
closely related to a conjectured duality between Wilson loops and scattering amplitudes,
which will be presented here.

\section{Gluon scattering amplitudes in $\mathcal{N}=4$ SYM}

\subsection{Perturbative results and BDS conjecture}

In order to state what the BDS conjecture implies it is useful to split a general planar $n$-point colour-ordered MHV amplitude $A_{n}$
into an infrared divergent part $D_{n\,{\rm IR}}$ and a finite part $F^{\rm (MHV)}_{n}$.
\begin{equation}\label{amplitudes}
\ln {A_{n}}/{A_{n}^{\rm tree}} = D_{n\,{\rm IR}} + F^{\rm (MHV)}_{n}(a,p_{i}\cdot p_{j}) + O(\epsilon_{\rm IR})\,.
\end{equation}
Here $D_{n\,{\rm IR}}$ contains poles in the infrared regulator $\epsilon_{\rm IR}$,
and as was mentioned in the introduction, it can be used to compute $\Gamma_{\rm cusp}$.
The 't Hooft coupling $a$ is related to the Yang-Mills coupling $g$ by $a = g^2 N / (8 \pi^2)$, and
$p_{i}^{\mu}$ are the $n$ light-like momenta of the scattering process. The structure of the IR divergent part 
is well-understood in gauge theory, see for example \cite{Dixon:2008gr} and references therein.
The BDS conjecture can be formulated as a statement about the finite part,
\begin{eqnarray} \label{BDS}
 F^{\rm (MHV)}_{n} &=& F^{\rm (BDS)}_{n}\,, \\
 F^{\rm (BDS)}_{n}(a,p_{i}\cdot p_{j}) &=& \frac{1}{2} \Gamma_{\rm cusp}(a)\, F^{\rm (MHV)}_{n;1}(p_{i}\cdot p_{j})\,. \nonumber
\end{eqnarray}
Note that the only coupling dependence on the r.h.s. of the second line of (\ref{BDS}) enters through the cusp anomalous dimension $\Gamma_{\rm cusp}(a)$. According to (\ref{BDS}), the functional dependence of $F^{\rm (MHV)}$ is coupling independent, 
and can therefore be determined for example by a one-loop computation. \footnote{$F^{\rm (MHV)}_{n;1}$ stands for the one-loop contribution to $F^{\rm (MHV)}_{n}$}
For example, the explicit functional form of $F^{\rm (BDS)}_{n}$ for $n=4$ is
\begin{eqnarray}\label{BDS4}
 F^{\rm (BDS)}_{4}(a,p_{i}\cdot p_{j}) &=& \frac{1}{4} \Gamma_{\rm cusp}(a) \left\lbrack \ln^2 \frac{s}{t} + {\rm const} \right\rbrack \,,\\
 F^{\rm (BDS)}_{5}(a,p_{i}\cdot p_{j}) &=&  \frac{1}{4} \Gamma_{\rm cusp}(a)  \left\lbrack \sum_{i=1}^{5} \ln \frac{s_{i,i+1}}{s_{i+1,i+2}} \ln \frac{s_{i+2,i+3}}{s_{i+3,i+4}} + {\rm const}  \right\rbrack  \,. \label{BDS5}
\end{eqnarray}
Here $s = (p_{1}+p_{2})^2$ and $t=(p_{2}+p_{3})^2$ are the usual Mandelstam variables, and similarly $s_{i,i+1}=(p_i + p_{i+1})^2$ are the 
kinematical invariants appearing in a five-particle scattering process.
The conjecture (\ref{BDS}) has been confirmed up to three loops for $n=4$ and two loops for $n=5$ gluons.
It seems very surprising that the functional form of $F^{\rm (MHV)}_{n}$ should be so simple, i.e. that the loop corrections to
$F^{\rm (MHV)}_{n}$ should take the simple form (\ref{BDS}). If the conjecture is true, one might expect some symmetry to
be responsible for this unexpected simplicity. We will see hints for such a symmetry by inspecting the 
integrals entering the loop corrections to $F^{\rm (MHV)}_{4}$. 

\subsection{Hints for a new symmetry}
Let us consider the one-loop corrections to $F^{\rm (MHV)}_{4}$. They are given by the following one-loop 
scalar box integral,
\begin{equation}\label{box-momenta}
I^{(1)} = \int\frac{d^{D} k }{k^2(k-p_1)^2(k-p_1-p_2)^2(k+p_4)^2}\,.
\end{equation}
In order to discover the new symmetry \cite{Broadhurst:1993ib,Drummond:2006rz}, one has to change variables to a {\it dual coordinate space} by
\begin{equation}
p_1=x_1-x_2\equiv x_{12}\,,\quad p_2=x_{23}\,,\quad p_3=x_{34}\,,\quad
p_4=x_{41}\,,\quad k=x_{15}\,,
\end{equation}
such that (\ref{box-momenta}) becomes
\begin{equation}\label{box-dual}
I^{(1)} = \int\frac{d^{D} x_5}{x_{15}^2 x_{25}^2 x_{35}^2 x_{45}^2}\,.
\end{equation}
For $D=4$ dimensions, $I^{(1)}$ in (\ref{box-dual}) is an integral familiar from the study of conformal correlation
functions. Indeed, it can be easily seen to be covariant under conformal transformations
in the dual coordinate space: since translation and rotation symmetry
are manifest, one only has to check covariance under dual conformal inversions,
\begin{equation}
x_i^\mu \to
x_i^\mu/x_i^2 \,,\qquad x^2_{ij}  \to \frac{x^2_{ij}}{x_i^2 x_j^2}\,,\qquad d^{D} x_{5} \to d^{D} x_{5} \, (x_{5}^2)^{-D}\,.
\end{equation}
Importantly, for $D=4$ the conformal weight at the integration point $x_{5}$ is exactly canceled 
between integration measure and the four 'propagators' connecting to the integration point.
Of course we cannot set $D=4$. The reason is that for on-shell momenta the distances $x_{i,i+1}^{\mu}$ are 
light-like, i.e. $x_{i,i+1}^2 = 0$, and this makes the integral $I^{(1)}$ infrared divergent in four dimensions.
From what was said before it is clear that if the momenta were off-shell, i.e. $x_{i,i+1}^2 \neq 0$, then $I^{(1)}$ would have an exact dual conformal
symmetry in four dimensions. We take this observation as a hint that there should be an underlying dual conformal symmetry, which is broken by infrared divergences. 
This expectation is further supported by the fact that the integrals corresponding to the higher loop corrections to the four-gluon amplitude also have this property,
at least up to four \cite{Bern:2006ew} and perhaps even to five loops \cite{Bern:2007ct}.\\

The dual conformal symmetry will become much more transparent and we will be able 
to make it more predictive within a new conjectured duality between scattering amplitudes
and Wilson loops, which will be described presently. 
As we will see, the Wilson loops naturally have a (broken) dual conformal symmetry. 
The latter implies (anomalous) dual conformal Ward identities for the Wilson loops, which 
can be used to make predictions for the scattering amplitudes.

\section{Duality between Wilson Loops and Scattering Amplitudes}

\begin{figure}
\psfrag{dots}[cc][cc]{$\bf\ldots$}
\psfrag{x1}[cc][cc]{$x_{1}$}
\psfrag{x2}[cc][cc]{$x_{2}$}
\psfrag{x3}[cc][cc]{$x_{3}$}
\psfrag{xn}[cc][cc]{$x_{n}$}
\psfrag{xnm}[cc][cc]{$x_{n-1}$}
\psfrag{p2}[cc][cc]{$p_{2}$}
\psfrag{p1}[cc][cc]{$p_{1}$}
\psfrag{pn}[cc][cc]{$p_{n}$}
\psfrag{pnm}[cc][cc]{$p_{n-1}$}
\centerline{{\epsfysize3.5 cm \epsfbox{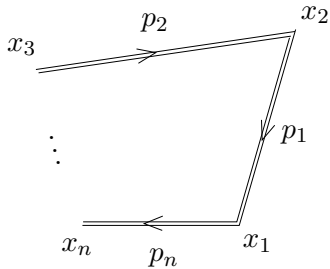}}} \caption[]{\small The integration contour $C_{n}$ of the Wilson loop $W(C_{n})$ dual to the $n$-gluon scattering amplitude.
The $p_{i}^{\mu}$ are the light-like momenta of the scattering process, related to the dual coordiantes $x^{\mu}_{i}$ by $x_i^\mu - x_{i+1}^\mu = p_i^\mu$.}
 \label{Fig-contour}
\end{figure}

A very interesting recent development in the AdS/CFT correspondence was an AdS prescription
for computing gluon scattering amplitudes at strong coupling \cite{Alday:2007hr}. It is presented in much more
detail in F. Alday's lectures given at this school. Interestingly, the AdS prescription of \cite{Alday:2007hr} suggests that a gluon scattering amplitude at strong
coupling is equivalent to the expectation value of a particular Wilson loop.
In the field theory, the relevant Wilson loop $W(C_{n})$ was first studied in this context in \cite{Drummond:2007aua}, and it is defined by
\begin{equation}
W(C_{n}) = \frac{1}{N} \, \langle 0| {\rm Tr}\, {\rm P}  \exp\left( i g  \oint_{C_{n}} dx_{\mu} A^{\mu} \right) | 0 \rangle \,.
\end{equation}
The gauge field $A^{\mu}$ is integrated along a closed contour $C_{n}$, which is depicted in Fig. \ref{Fig-contour}. 
It is a polygon whose corners are coordinates $x^{\mu}_{i}$
in a dual coordinate space related to the gluon momenta by \footnote{Here and in the following $x_{i+n}\equiv x_{i}$ is tacitly implied for the $n$-cusp Wilson loop.}
\begin{equation}\label{dualcoords2}
x_i^\mu - x_{i+1}^\mu := p_i^\mu\,.
\end{equation}
Interestingly, this is precisely the relation between gluon momenta and dual coordinates used to
study the dual conformal properties of the scalar integrals in the previous section.\\

The general structure of the Wilson loops is very similar to that of the scattering amplitudes, c.f. equation (\ref{amplitudes}):
\begin{equation}
\ln W(C_{n}) = D_{n\,{\rm UV}} \; + F_n^{\rm (WL)}(a,x_{ij}^2) + O(\epsilon_{\rm UV})\,.
\end{equation}
Here $D_{n\,{\rm UV}}$ contains ultraviolet poles associated with the cusps of the Wilson loop (for more details see \cite{Korchemskaya:1992je} and
references therein).
According to the conjectured duality,
\begin{equation}\label{duality}
F^{\rm (MHV)}_{n} = F^{\rm (WL)}_{n} + {\rm const} + O(1/N)\,,
\end{equation}
to all orders in the coupling constant $a$.
More precisely, the duality relation (\ref{duality}) states that, upon identification of the gluon momenta with the dual coordiantes according to (\ref{dualcoords2}), 
the finite part of the MHV scattering amplitude should coincide with the finite part of the Wilson loop, up to a constant and up to non-planar corrections.

\subsection{Tests of the duality}

\begin{figure}\label{Fig-WL2}
{\includegraphics[height = 30mm]{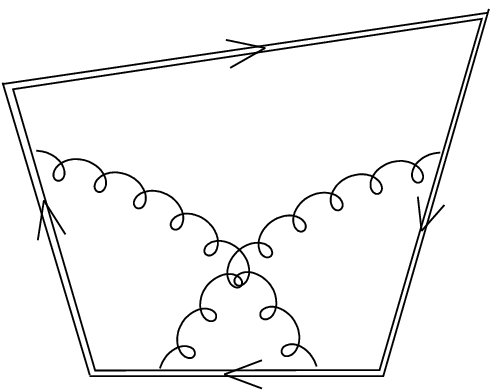}}\hspace{0.5cm}{\includegraphics[height = 30mm]{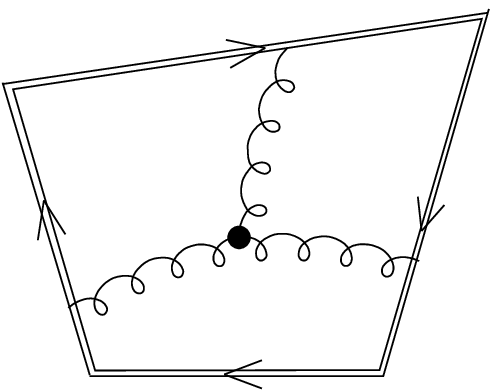}}\hspace{0.5cm}{\includegraphics[height = 30mm]{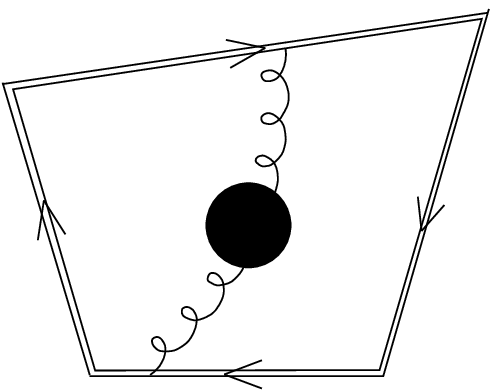}}
\caption{Three representative diagrams contributing to the expectation value of the four-cusp Wilson loop at two loops. The blob denotes a one-loop propagator correction.}
\end{figure}

It was shown in \cite{Drummond:2007aua} that the duality relation (\ref{duality}) holds true at one loop and $n=4$ points.
This was extended to arbitrary $n$ at one loop in \cite{Brandhuber:2007yx}.
At one loop, the computation of the Wilson loop entering the duality involves integrating
a free gluon propagator along the polygonal contour $C_{n}$. It is clear that such a computation is
insensitive to the specific details of $\mathcal{N}=4$ SYM such as e.g. interaction vertices and field content.
For this reason it seems very important to investigate the validity of the duality to higher orders
in perturbation theory. \\

Therefore, a two-loop computation of the Wilson loop for $n=4$ and $n=5$ was carried out in \cite{Drummond:2007cf,Drummond:2007au}.
Some representative Feynman graphs are depicted in Figure \ref{Fig-WL2}.
After some remarkable cancelations, the result indeed reduces to the functional form of the BDS ansatz,
written in the dual coordinates, and hence the duality holds at two loops for $n=4$ and $n=5$.\\

Just as for the scattering amplitudes, it may seem surprising that the loop corrections to Wilson
loops should take the simple form (\ref{BDS}). 
We will see presently that this simplicity, at least for $n=4$ and $n=5$, is a consequence of dual conformal symmetry. 

\subsection{(Broken) conformal Ward identities for light-like Wilson loops}

In contrast to the scattering amplitudes, which are defined in momentum space, the Wilson loops are defined in configuration space
(which is dual from the point of view of the scattering amplitudes). Therefore we can directly exploit the conformal symmetry of $\mathcal{N}=4$ SYM 
which acts in configuration space. A crucial observation is that the contour on which the Wilson loop is defined is stable under
conformal transformations: under the latter, a light-like polygon is mapped into another light-like polygon.
This and the conformal invariance of the action of $\mathcal{N}=4$ SYM allow us to derive conformal
Ward identities for the Wilson loops.\\

A very important effect arises due to the UV divergences of the Wilson loops. The dimensional
regulator breaks the conformal symmetry, which leads to an anomalous term in the Ward identity.
We stress that in order to be able to make quantitative predictions for the Wilson loops, it
is crucial to control this anomalous contribution.
The conformal boost Ward identity, first proposed in \cite{Drummond:2007cf} and
then proven in \cite{Drummond:2007au}, reads
\begin{equation}\label{CWI}
\sum_{i=1}^n  \left[ 2 x_i^\mu (x_i \cdot
\partial_{x_i}) - x_i^2
\partial_{x_i}^\mu \right] { F^{\rm (WL)}_n} = \frac{1}{2} { \Gamma_{\rm cusp}(a)}\sum_{i=1}^n x_{i,i+1}^\mu \ln \Bigl(
\frac{x_{i,i+2}^2}{x_{i-1,i+1}^2} \Bigr)\,.
\end{equation}
Note that the anomalous term on the right-hand side of (\ref{CWI}) is coupling-dependent,
but only through the cusp anomalous dimension $\Gamma_{\rm cusp}(a)$.\\

It turns out that (\ref{CWI}) has very strong implications. For $n=4$ and $n=5$ points, it completely
fixes the functional form of the Wilson loop, namely
\begin{eqnarray}\label{CWI-sol4}
F^{\rm (WL)}_{4} &=&   \frac{1}{4} { \Gamma_{\rm
cusp}(a)}\, { \ln^2\Bigl(\frac{x_{13}^2}{x_{24}^2}\Bigr)} + \text{
 const } \,, \\
F^{\rm (WL)}_{5} &=&  \frac{1}{4} { \Gamma_{\rm cusp}(a)} \sum_{i=1}^5 \ln
 \Bigl(\frac{x_{i,i+2}^2}{x_{i+1,i+3}^2}\Bigr) \ln
 \Bigl(\frac{x_{i+2,i+4}^2}{x_{i+3,i}^2}\Bigr) + \text{ const } \label{CWI-sol5}\,.
\end{eqnarray}
We see that equations (\ref{CWI-sol4}) and  (\ref{CWI-sol5}) correspond precisely to the BDS formula for scattering amplitudes, c.f. equations (\ref{BDS4}) and (\ref{BDS5}),
rewritten in the dual coordinates.
This all-order result allows us to draw the following conclusions:
\begin{itemize}
  \item for $n=4$, it confirms the duality (\ref{duality}) to three loops, since the BDS formula for gluon scattering amplitudes holds in this case \cite{Bern:2005iz}. 
It also agrees with the result obtained in \cite{Alday:2007hr} at strong coupling using the AdS/CFT correspondence;
  \item if one assumes the duality between scattering amplitudes and Wilson loops, the conformal Ward identity
  for Wilson loops explains why the BDS ansatz is true for $n=4,5$ points.
\end{itemize}

Starting from $n=6$ points, a new feature appears: one can build conformal invariants which take the form of cross-ratios \footnote{Usually, cross-ratios can already be built at four points. Here the conditions $x_{i,i+1}^2=0$ postpone the appearance of conformal cross-ratios until six points.},
\begin{equation}
\mathbb{K}^{\mu}\, \frac{x_{ij}^2 x^2_{kl}}{x^2_{ik}x^2_{jl}} = \sum_{m=1}^n  \left[ 2 x_m^\mu (x_m \cdot
\partial_{x_m}) - x_m^2
\partial_{x_m}^\mu \right]\, \frac{x_{ij}^2 x^2_{kl}}{x^2_{ik}x^2_{jl}} = 0 \,.
\end{equation}
At six points, there are three such invariants,
\begin{equation}
u_1 = \frac{x_{13}^2 x_{46}^2}{x_{14}^2 x_{36}^2}, \qquad u_2 = \frac{x_{24}^2
x_{15}^2}{x_{25}^2 x_{14}^2}, \qquad u_3 = \frac{x_{35}^2 x_{26}^2}{x_{36}^2
x_{25}^2}\,.
\end{equation}
Hence for general $n\ge6$, a particular solution of (\ref{CWI}) is still given by the BDS ansatz, but one can
 always add an arbitrary function of conformal invariants to it. For example, at six points we have
\begin{equation}\label{CWI-sol6}
F^{\rm (WL)}_{6} = F^{\rm (BDS)}_{6} + f(a;u_{1},u_{2},u_{3})\,.
\end{equation}
Dual conformal symmetry does not restrain the function $f(a;u_{1},u_{2},u_{3})$,
and therefore it seems very interesting to ask whether the latter receives non-trivial loop corrections,
and whether the duality (\ref{duality}) holds for $n=6$ at two loops.

\subsection{Beyond dual conformal symmetry: six-gluon amplitude}

In order to shed light on these questions, a two-loop computation of the hexagonal Wilson loop was performed
in \cite{Drummond:2007bm}. 
It was found that, in perfect agreement with the Ward identity (\ref{CWI}), $F^{\rm (WL)}_{6}$ is
correctly described by (\ref{CWI-sol6}), however with a non-trivial (non-constant) function $f(a;u_{1},u_{2},u_{3})$ at
two loops. Therefore, the hexagonal Wilson loop at six points is not given by the BDS ansatz for the scattering amplitudes.
Since the corresponding six-gluon amplitude had not been computed at this point, this meant that 
either the duality with scattering amplitudes or the BDS ansatz had to fail.
Indications that the BDS ansatz should break down came from \cite{Alday:2007he} and \cite{Bartels:2008ce}.
It should be stressed that a breakdown of the BDS ansatz does not automatically mean that the duality is true,
because both of them could break down at the same time.\\

Very recently, these questions could be answered when the calculation of the two-loop six-gluon MHV amplitude was completed \cite{Bern:2008ap}. The authors of \cite{Bern:2008ap} found
that the BDS ansatz needs to be corrected. Moreover, a numerical comparison between the result for the hexagonal Wilson loop
and the six-gluon amplitude was carried out \cite{Drummond:2007bm,Bern:2008ap}, and it was found that within the numerical accuracy the duality holds! 
Given this further evidence in favour of the duality (\ref{duality}) we are 
confident that it should hold in general.

\section{Conclusions and outlook}

We presented evidence for a new duality between gluon scattering amplitudes 
and Wilson loops in $\mathcal{N}=4$ SYM. Several two-loop calculations were undertaken and 
the results agreed with the duality. Moreover, it was shown that the Wilson loops 
have to obey an all-order conformal Ward identity.
If the duality is true, then the Ward identity explains why the BDS ansatz for
gluon scattering amplitudes holds for $n=4$ and $n=5$ gluons. For $n = 6$ and two
loops, the BDS ansatz is incorrect and has to be modified by a function of dual conformal invariants,
in complete agreement with the duality and the dual conformal Ward identity.\\

The results described in this talk were limited to MHV scattering amplitudes,
which correspond to the simplest possible helicity configurations. There
are many other helicity configurations, and it is natural to ask whether the duality
can be extended to these as well, and whether one can find a dual conformal 
symmetry in non-MHV amplitudes.\\

Shortly after this talk was given, the second question was answered positively.
It was discovered \cite{dhks5} that the dual conformal symmetry described in this talk can
be extended to a dual superconformal symmetry. Moreover, the dual superconformal
symmetry is a property of all amplitudes, with MHV and non-MHV helicity configurations.
Hints that scattering amplitudes should have a dual superconformal symmetry were
also found using the AdS/CFT correspondence \cite{Berkovits:2008ic,Beisert:2008iq}. It would be extremely
interesting if one could learn from the AdS/CFT correspondence how to extend the 
duality to non-MHV amplitudes.
Newer references discussing the dual superconformal symmetry include \cite{Brandhuber:2008pf,dhks6,Drummond:2008cr}.

\section{Acknowledgements}
I am grateful to the organisers of the 48th Cracow School of Theoretical Physics for the opportunity to give this talk.
It is a pleasure to thank J.~M.~Drummond, G.~P.~Korchemsky and E.~Sokatchev for collaboration on the topics presented here.
I thank G.~P.~Korchemsky for his comments on the draft of the manuscript.
This research was supported in part by the French Agence Nationale 
de la Recherche under grant ANR-06-BLAN-0142.


\end{document}